\documentclass[12pt,preprint]{aastex}
\usepackage[dvips]{color}
\usepackage{graphicx}

\def\msol{\hbox{\kern 0.20em $M_\odot$}}
\def\lsol{\hbox{\kern 0.20em $L_\odot$}}
\def\rsol{\hbox{\kern 0.20em $R_\odot$}}
\def\sr{\hbox{\kern 0.20em sr}}
\def\srmu{\hbox{\kern 0.20em sr$^{-1}$}}
 
\def\g{\hbox{\kern 0.20em g}}
\def\gmu{\hbox{\kern 0.20em g$^{-1}$}}
\def\kg{\hbox{\kern 0.20em kg}}
\def\pc{\hbox{\kern 0.20em pc}}
 
\def\mum{\hbox{\kern 0.20em $\mu$m}}
\def\mumd{\hbox{\kern 0.20em $\mu$m$^{-2}$}}
\def\cm{\hbox{\kern 0.20em cm}}
\def\m{\hbox{\kern 0.20em m}}
\def\km{\hbox{\kern 0.20em km}}
\def\nm{\hbox{\kern 0.20em nm}}
 
\def\s{\hbox{\kern 0.20em s}}
\def\h{\hbox{\kern 0.20em h}}
\def\sec{\hbox{\kern 0.20em sec}}
\def\min{\hbox {\kern 0.20em min}}
\def\smu{\hbox{\kern 0.20em s$^{-1}$}}
\def\smd{\hbox{\kern 0.20em s$^{-2}$}}
\def\an{\hbox{\kern 0.20em an}}
\def\anmu{\hbox{\kern 0.20em an$^{-1}$}}
\def\deg{\hbox{\kern 0.20em $^{\rm o}$}}
\def\yr{\hbox{\kern 0.20em yr}}
\def\yrmu{\hbox{\kern 0.20em yr$^{-1}$}}
\def\Myr{\hbox{\kern 0.20em Myr}}
\def\Mymu{\hbox{\kern 0.20em Myr$^{-1}$}}
\def\K{\hbox{\kern 0.20em K}}
\def\pcmu{\hbox{\kern 0.20em pc$^{-1}$}}
\def\pcmd{\hbox{\kern 0.20em pc$^{-2}$}}
\def\pcmt{\hbox{\kern 0.20em pc$^{-3}$}}
\def\kms{\hbox{\kern 0.20em km\kern 0.20em s$^{-1}$}}
\def\kmpd{\hbox{\kern 0.20em km$^{2}$}}
\def\kpc{\hbox{\kern 0.20em kpc}}
\def\cms{\hbox{\kern 0.20em cm\kern 0.20em s$^{-1}$}}
\def\erg{\hbox{\kern 0.20em erg}}
\def\ergs{\hbox{\kern 0.20em erg}}
\def\cmpd{\hbox{\kern 0.20em cm$^2$}}
\def\cmmd{\hbox{\kern 0.20em cm$^{-2}$}}
\def\cmms{\hbox{\kern 0.20em cm$^{-6}$}}
\def\cmpt{\hbox{\kern 0.20em cm$^3$}}
\def\cmmt{\hbox{\kern 0.20em cm$^{-3}$}}
\def\mpd{\hbox{\kern 0.20em m$^2$}}
\def\mmd{\hbox{\kern 0.20em m$^{-2}$}}
\def\mpt{\hbox{\kern 0.20em m$^3$}}
\def\mmt{\hbox{\kern 0.20em m$^{-3}$}}
\def\mujy{\hbox{\kern 0.20em $\mu$Jy}}
\def\mjy{\hbox{\kern 0.20em mJy}}
\def\Mj{\hbox{\kern 0.20em MJy}}
\def\jy{\hbox{\kern 0.20em Jy}}
\def\ghz{\hbox{\kern 0.20em GHz}}
\def\srmd{\hbox{\kern 0.20em sr$^{-1}$}}
\def \kms{km~$\rm{s}^{-1}$}

\def \mum{$\mu$m}

\def\G{\hbox{\kern 0.20em G}}

\def\h13cop{\hbox{H$^{13}$CO$^{+}$}}

\def\S+{\hbox{S{\small II}}}

\slugcomment{The Evolving ISM in the Milky Way \& Nearby Galaxies}

\shortauthors{Tasker et al.}

\begin{document}

\newcommand{\jfourteen}{\hbox{$J=14\rightarrow 13$}}
 \title{Simulating the ISM in Global Models of Disk Galaxies}

\author{Elizabeth J. Tasker\altaffilmark{1},
Greg L. Bryan\altaffilmark{2},
Jonathan C. Tan\altaffilmark{1}}

\altaffiltext{1}{Department of Astronomy, University of Florida, Gainesville, FL 32611, USA.}
\altaffiltext{2}{Department of Astronomy, Columbia University, New York, NY 10027, USA.}

\begin{abstract}
Until recently, simulations that modeled entire galaxies were
restricted to an isothermal or fixed 2- or 3-phase interstellar medium
(ISM). This obscured the full role of the ISM in shaping the observed
galactic-scale star formation relations. In particular the Kennicutt
relation suggests that star formation rates depend in a simple way on
global galactic quantities, such as mean gas mass surface density and
dynamical time. Contrary to this, observations of nearby star-forming
regions, including images from the Spitzer telescope, show that all
the way down to ``local-scales'' ($\sim$ parsecs) star formation is a
highly clustered process with the gas existing at a wide range of
densities and multiple phases, and seemingly decoupled from the
larger-scale Galaxy. Many different physical processes appear to be
influencing the star formation rate, including heating and cooling of
the gas, ``turbulent'' energy injection from outflows, winds and
supernova blastwaves, ionization, and magnetic field support. Here we
present results from global simulations of disk galaxy star formation
that include a fully multiphase ISM and compare to simpler isothermal
models. We also investigate the effect of background heating. We
discuss the process of giant molecular cloud formation as one of the
possible links between the galactic and local scales of star formation.
\end{abstract}


\lefthead{Elizabeth J. Tasker et al.}

\righthead{Simulating the ISM in Global Disk Galaxies}

\section{Introduction}

Lack of a detailed understanding of the star formation process is one
of the main factors limiting theoretical and numerical models of
galaxy formation and evolution. The relatively simple gas dynamics and
star formation relations exhibited by galaxies on their global
($>$kpc) scale (e.g. Kennicutt 1998), mask the true complexity that is
revealed from close examination of star-forming regions on their
``local'' scale, $\sim$parsecs, typically inside giant molecular
clouds (GMCs). Here it is suspected that star formation is a result of a
complex interplay and competition between, on the one hand, self-gravity and
compressional processes such as shocks and, on the other hand, various
processes that help support the gas by generating thermal, radiation,
magnetic, ``turbulent'' (i.e. due to nonthermal motions) and cosmic
ray pressures. Many of these latter processes are driven by the newly-formed
stars, through their protostellar outflows, winds, radiation, and supernovae.

Star formation is a highly clustered process (e.g. Lada \& Lada 2003),
and the gas that is forming stars at high efficiency, i.e. in
star-forming clumps that are the progenitors of star clusters, is
typically at a much higher pressure than the ambient galactic scale
ISM: it is effectively decoupled. This is also illustrated by the fact
that only a small fraction of the total gas mass in the star-forming
part of a galaxy is actively forming stars. Thus the observed
correlations between total gas mass surface density, $\Sigma_{\rm
gas}$, and star formation rate surface density, $\Sigma_{\rm SFR}$ are
somewhat mysterious. Using disk averages, Kennicutt (1998) finds
$\Sigma_{\rm SFR}\propto \Sigma_{\rm gas}^{1.4}$ in a sample of normal
spiral galaxies and circumnuclear starbursts. 
An equally good description is $\Sigma_{\rm SFR} \propto \Sigma_{\rm gas}/t_{\rm
orb}$, where $t_{\rm orb}$ is the orbital time (at the outer radius of
the star-forming disk). More local ($\sim$kpc scale) regions of
galaxies follow similar average relations, but with a wider scatter
(e.g. Kennicutt et al. 2007). To understand the connection between
global star formation relations and the scale of structures actually
involved in star formation, it seems clear that we need to understand
the process of dense, cold cloud formation from the diffuse ISM, which
requires resolution of the multiphase nature of the gas.

Until recently, simulations have struggled to connect global galactic
dynamics with a detailed multiphase treatment of the ISM.
The difficulty of resolving parsec-scale physics over tens of
kiloparsecs has meant that previous simulations have focused either on
a small volume of the galaxy where the ISM can be properly modeled
\citep{Slyz2005}, or simulated the whole galaxy at the cost of an
isothermal or fixed 2 or 3-phase temperature ISM \citep{Li2005,
Robertson2004}. However, more recent work has begun to bridge this gap
and include a self-consistent multiphase ISM in global galaxy models
\citep{Tasker2006, Tasker2008, Wada2007}. By leading to a better
understanding of the physical processes that control galactic star
formation rates, these simulations can also yield insights into the
uncertainties involved in using simplified models of the
thermodynamics and phase structure of the ISM in larger cosmological
simulations.

Here, we present global galaxy disk simulations with different ISM
properties and compare their gas structure and star formation
rates. We also present preliminary results on dense cloud formation,
especially with regards to cloud rotation.

\section{Global Galaxy Models}
\begin{figure}[!t]
\centerline{
\includegraphics[width=\textwidth]{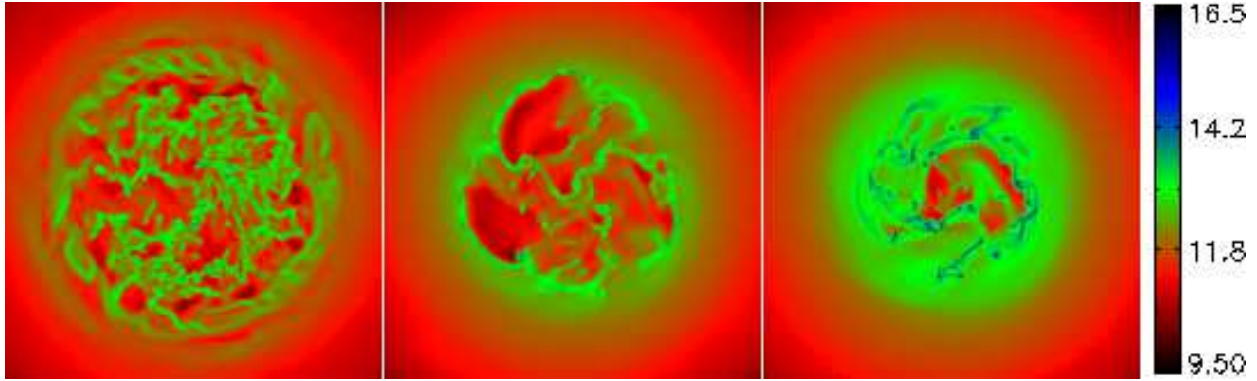}
}
\caption{\label{fig:proj} Gas mass surface density, $\Sigma_{\rm
gas}$, at 377\,Myrs for (left to right); ISM \#1 (cooling only), ISM
\#2 (background heating term included) and ISM \# 3 (isothermal gas),
all without feedback. Scale is to the base-10 logarithm with units of
M$_\odot$Mpc$^{-2}$ and images are 60\,kpc across.}
\end{figure}

To explore the effects of including a self-consistent multiphase ISM
in global galaxy models, we compared three galaxy simulations with
different ISM properties (see Tasker \& Bryan 2008 for more
details). In the first model (ISM \#1), the gas was allowed to cool
radiatively.
In our second model (ISM \#2), an additional fixed background heating
term was included to represent photoelectric heating.
In our final model (ISM \#3), the gas temperature was fixed at
$10^4$\,K. 
The effects of including feedback from Type II supernovae via thermal energy injection 
were also explore in models \#1 and \#2.

The simulations were run using the 3D hydrodynamics code,
\emph{Enzo}, which is an adaptive mesh refinement code
well suited to modeling media where a range of temperatures and
densities are present.

The initial set-up for each simulation is an isolated, pure gas
Milky Way-sized disk, sitting in a static NFW dark matter
potential. The density profile matched that of the luminosity of a disk galaxy; exponential in the radial direction and $sech^2$ height profile. 

The disk initially fragments through gravitational
instability and forms stars. The results presented here are taken
377\,Myrs after the start of the simulation, when the disk has
settled down to an approximately constant star formation rate.

Figure~\ref{fig:proj} shows the gas mass surface density, $\Sigma_{\rm gas}$, for the three
different ISM types. The difference in the gas structure is very
noticeable, but all exhibit fragmentation out to a well-defined
radius. Beyond this radius, there is still plenty of gas, but it is
gravitationally stable and does not form stars. In the case where we
include background heating (ISM \#2), the disk fragments out to a
smaller radius than in ISM \#1, because the heating provides an extra
pressure against the fragmentation. The structure of the fragmented
region also differs significantly, showing large voids of low density,
hotter gas. In ISM \# 3, where the temperature is fixed, a different
structure is seen again. The increased temperature in the disk results
in a much higher Jeans length for gravitational collapse, producing
larger dense gas clouds than in either of the other two models. 

\begin{figure}[!t]
\centerline{
\includegraphics[angle=-90,width=\textwidth]{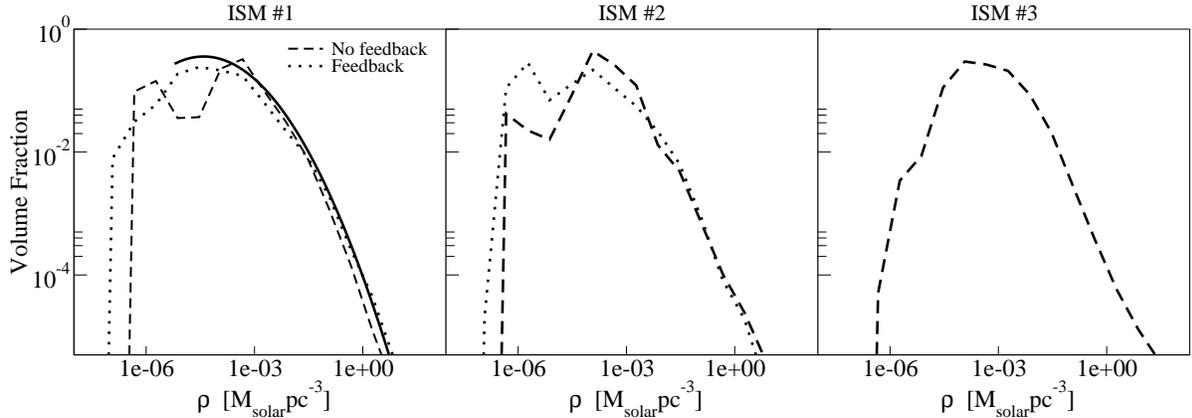}
}
\caption{\label{fig:pdfs} PDF of the volume weighted gas density at 377\,Myrs. A lognormal fit is shown as a solid line in the first panel.}
\end{figure}

A more quantitative view of the ISM of these galaxies is shown via the
probability density functions (PDFs) of gas density (weighted by
volume) in Figure~\ref{fig:pdfs}. The first two ISM models are shown
for runs with and without feedback from Type II SNe, which is
implemented via localized heating of gas around star cluster particles
in the simulation. What is obvious from this figure is that, while the
medium and low density gas show significant differences from plot to
plot, the high density end takes a far more uniform shape. In the left
panel, showing results from ISM \#1, the solid line picks out a
lognormal profile that is a reasonable fit for the high density end of
the distributions. These results show that SN feedback has relatively
little impact on the volume fraction of dense gas. This volume
fraction is also fairly invariant between the different ISM models
considered here. This may help to explain the universality of the
Kennicutt relation in different galactic environments, if the star
formation rate from dense gas scales as the mass of dense gas divided
by the local galactic dynamical time (see \citet{Elmegreen2002} for a further
discussion). One possible physical mechanism that can produce such a
scaling is the triggering of active star-forming regions by GMC-GMC
collisions \citep{Tan2000}.

\begin{figure}[!ht]
\centerline{
\includegraphics[width=9cm]{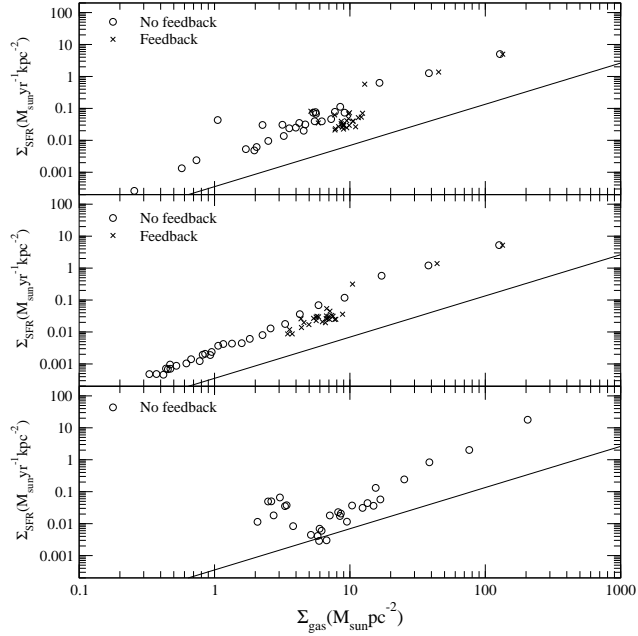}}
\caption{The globally-averaged surface density of star formation versus gas mass surface density for the ISM models 1 (top), 2 (middle), and 3 (bottom). Each data point represents a different time in simulations, which all involve a declining star formation rate as gas is depleted from the galaxies. The solid line shows the observational result of Kennicutt (1989).}
\label{fig:sk}
\end{figure}

Figure~\ref{fig:sk} shows the Kennicutt relation of globally-averaged
$\Sigma_{\rm SFR}$ vs. $\Sigma_{\rm gas}$ for all galaxy models. The
observed gradient is reproduced well in both ISM \#1 and ISM \#2, but
less well in the isothermal case. Despite this success, the star
formation rate is over-estimated by every model. The inclusion of
feedback does reduce the discrepancy, but only by a small factor.
However, because the minimum cell size in these simulations is
typically 25-50~pc, there is considerable freedom to choose the
normalization of the star formation efficiency in the subgrid
model. In the simulations shown here, Tasker \& Bryan (2008) adopted a
formation efficiency of 5\% per dynamical time in cells that met
certain criteria, namely a density above a critical density, a gas
mass greater than the Jeans mass, convergent gas flow into the cell,
and a cooling time less than the dynamical time (for non-isothermal
models). The critical density adopted for these simulations was small:
$n=0.02\:{\rm cm^{-3}}=7\times 10^{-4}\:M_\odot\:{\rm yr^{-1}}$, so
that a large volume fraction of the gas could be involved in star
formation. Improved star formation subgrid models are being developed
to more closely match observed star formation efficiencies, and these
will likely lead to a lower normalization of the simulated Kennicutt
relation.

\section{Dense Cloud Formation and Cloud Rotation}

\begin{figure}[!ht]
\centerline{
\includegraphics[width=5cm]{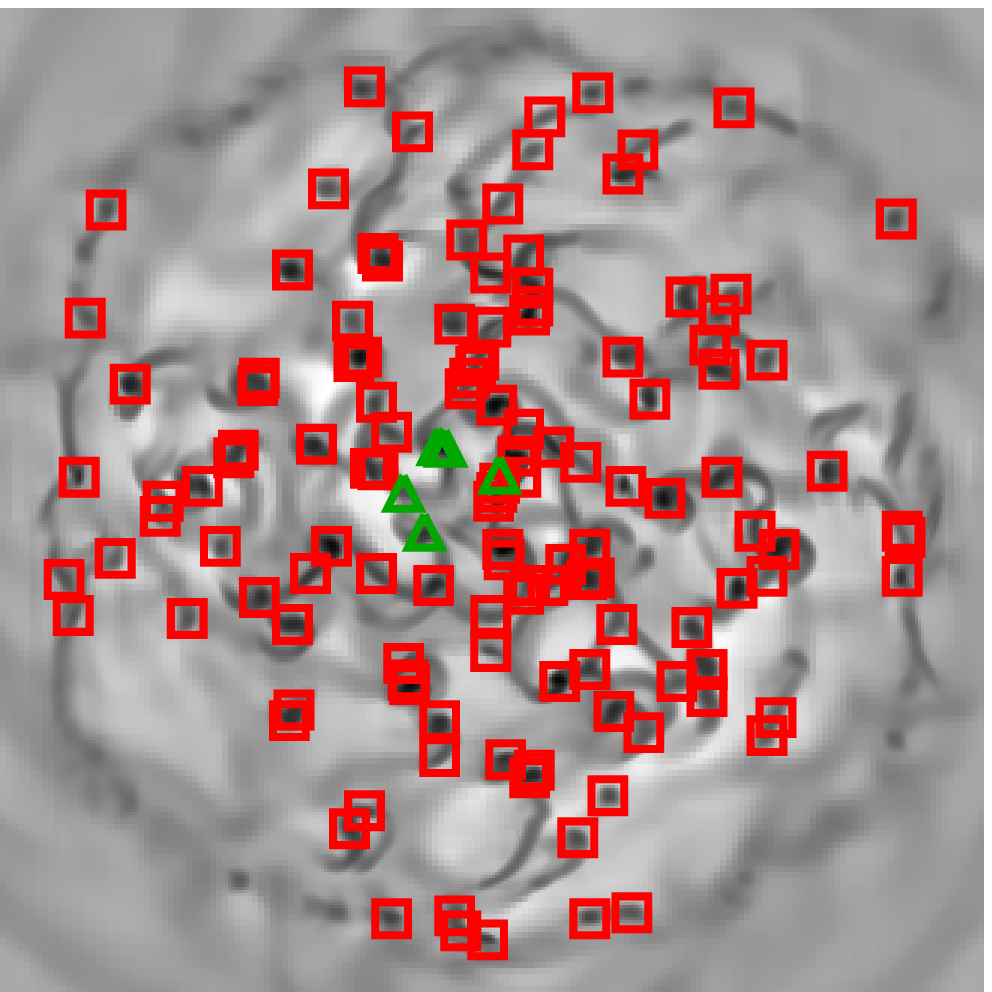}
\includegraphics[width=5cm]{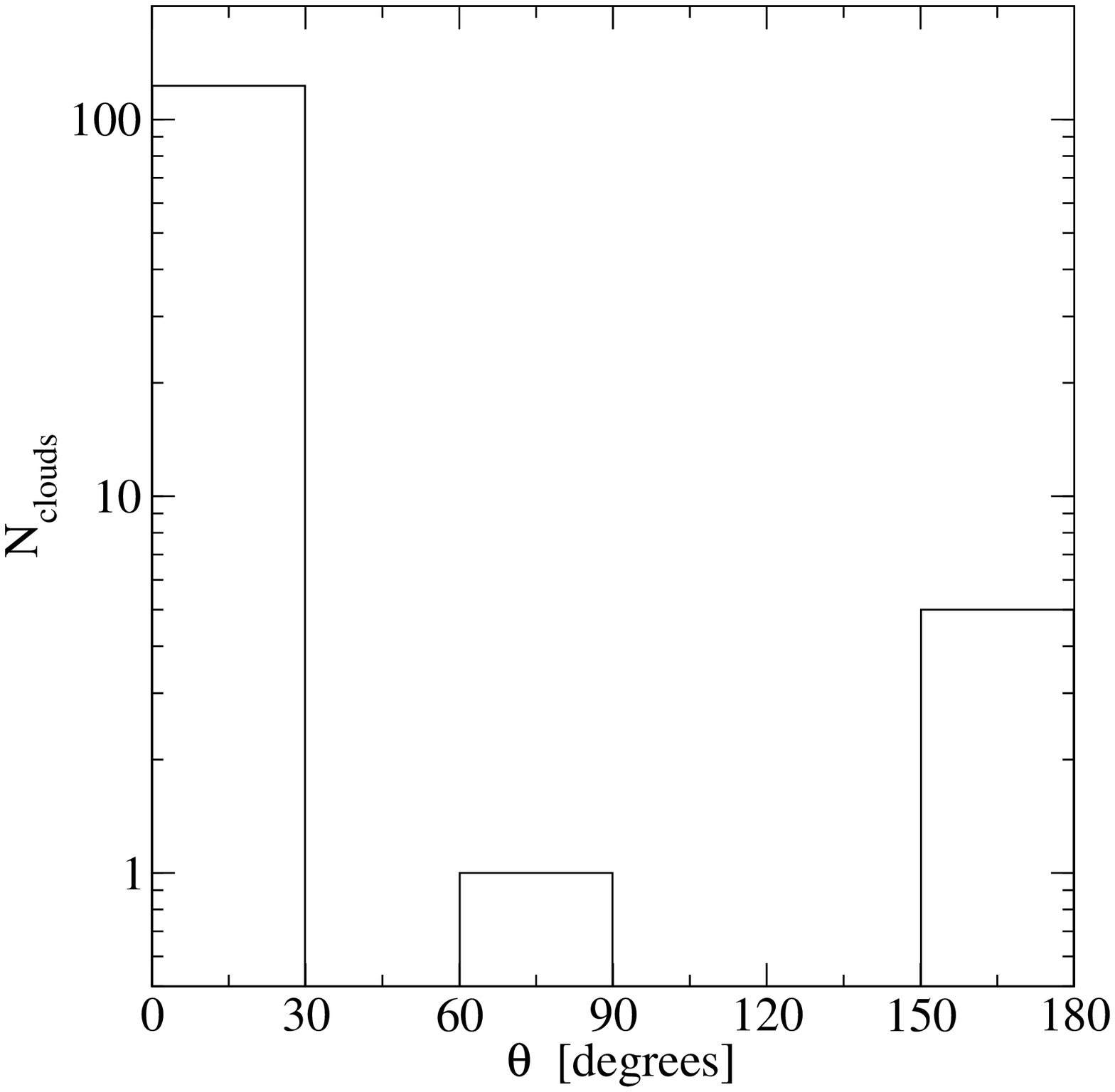}
}
\centerline{
\includegraphics[width=5cm]{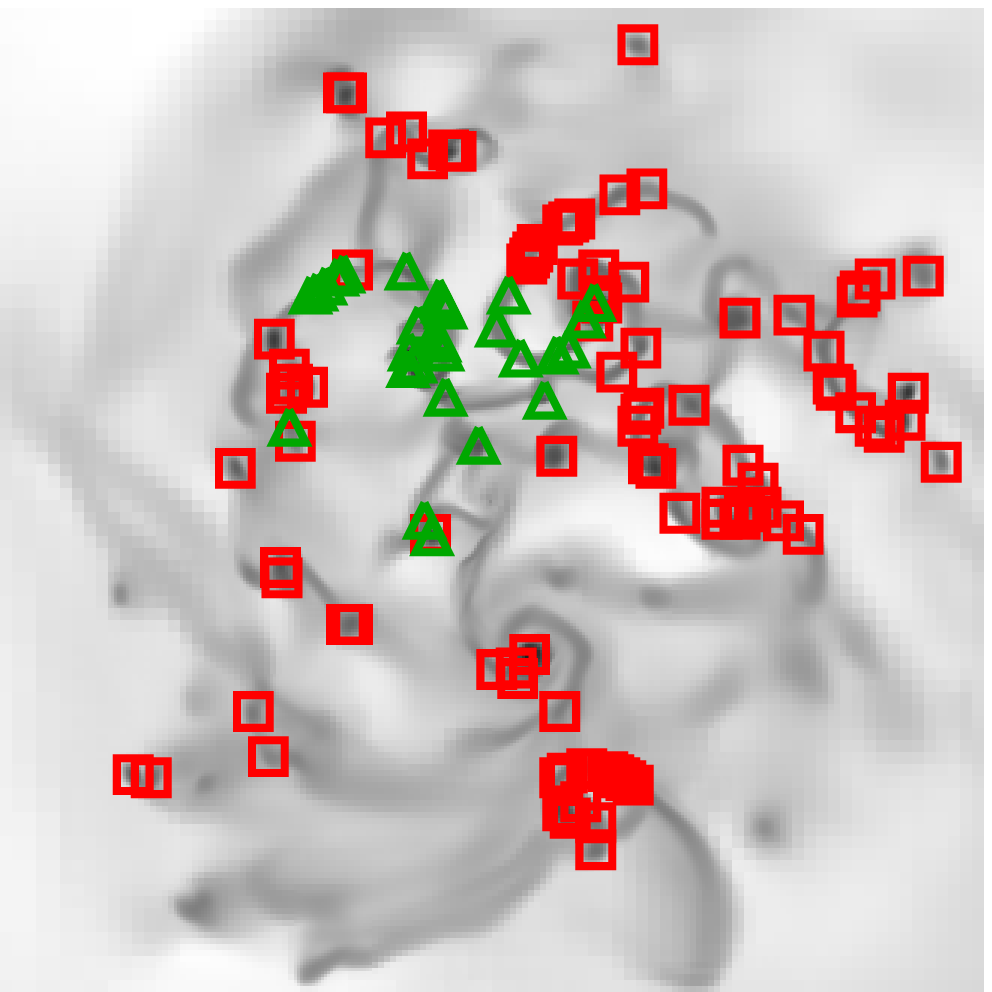}
\includegraphics[width=5cm]{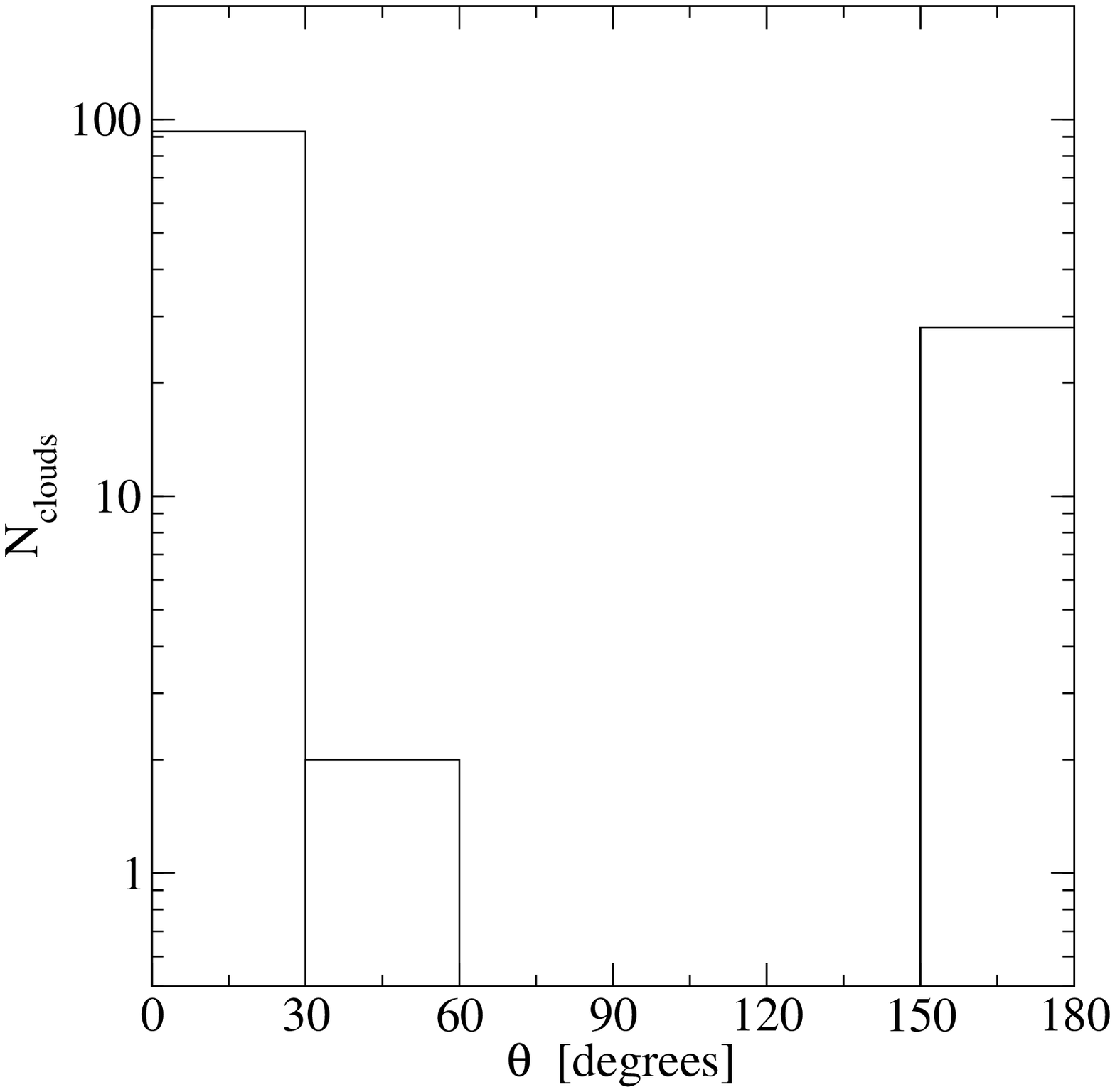}
}
\caption{Top left: gas mass surface density projection of a galaxy model at 189 Myrs after the start of a simulation with no star formation or feedback. ISM type matches that of ISM\,\#1 in the previous section. Red squares represent clouds identified with prograde motion with respect to the galaxy, green triangles rotate retrograde. Top right: Distribution of cloud rotation axes with respect to that of the galaxy ($0^\circ$ is prograde, $180^\circ$ is retrograde). Bottom left and right: same plots for the disk 566\,Myrs after the start of the simulation, where a significant fraction, $\sim 1/3$, of the clouds have now become retrograde, especially in the more dynamically-evolved innner regions.}
\label{fig:clouds}
\end{figure}

The ability to model the structure of the ISM on global scales opens a
new door in the study of star formation. Our highest resolution
simulations are now at the level of the largest GMCs, in which the
majority of the cold, star-forming gas mass is tied up. Following the
evolution of these clouds therefore enables us to track the earliest
stages of star formation and see how it is affected by global
properties such as galactic potential and feedback.

Figure~\ref{fig:clouds} show the results from early, low-resolution simulations, which
imply that the GMCs form with a prograde rotation \citep[see also][]{Kim2003} but then
interact and merge to generate a broader distribution of rotation
directions, including a substantial fraction of retrograde rotators.
Future simulations will quantify this evolution and, from comparison
to observed clouds \citep[e.g.][]{Rosolowsky2003}, place constraints on
GMC lifetimes.

\vspace{-10pt}
\acknowledgements
EJT acknowledges support from a Theoretical Astrophysics Postdoctoral
Fellowship from Dept. of Astronomy/CLAS, University of Florida and
thanks the University of Florida High-Performance Computing Center for
providing computational resources and support. JCT acknowledges
support from NSF CAREER grant AST-0645412.

\end{document}